\documentclass[sigconf,natbib=false]{acmart}
\AtBeginDocument{%
  \providecommand\BibTeX{{%
    \normalfont B\kern-0.5em{\scshape i\kern-0.25em b}\kern-0.8em\TeX}}}

\usepackage{tabularx}

\usepackage{fancyhdr}
\usepackage{algpseudocode}
\usepackage{algorithm}

\usepackage[style=ACM-Reference-Format,backend=biber,sorting=none,style=acmnumeric,giveninits=true]{biblatex}
\copyrightyear{2023} 
\acmYear{2023} 
\setcopyright{rightsretained} 
\acmConference[SC '23]{The International Conference for High Performance Computing, Networking, Storage and Analysis}{November 12--17, 2023}{Denver, CO, USA}
\acmBooktitle{The International Conference for High Performance Computing, Networking, Storage and Analysis (SC '23), November 12--17, 2023, Denver, CO, USA}
\acmDOI{10.1145/3581784.3627040}
\acmISBN{979-8-4007-0109-2/23/11}

\begin{document}

\title{Towards Exascale Computation for Turbomachinery Flows}

\author{Yuhang Fu}
\authornote{Both authors contributed equally to this research.}
\email{121fuyh@zju.edu.cn}
\orcid{0009-0006-6686-2414}
\author{Weiqi Shen}
\authornotemark[1]
\email{weiqishen1994@gmail.com}
\orcid{0000-0001-6203-0182}
\affiliation{%
  \institution{Zhejiang University}
  \city{Hangzhou}
  \country{China}
}

\author{Jiahuan Cui}
\authornote{Corresponding authors}
\email{jiahuancui@intl.zju.edu.cn}
\orcid{0000-0002-4962-3052}
\author{Yao Zheng}
\authornotemark[2]
\email{yao.zheng@zju.edu.cn}
\orcid{0000-0003-2126-2897}
\affiliation{%
  \institution{Zhejiang University}
  \city{Hangzhou}
  \country{China}
}

\author{Guangwen Yang}
\authornotemark[2]
\email{ygw@tsinghua.edu.cn}
\orcid{0000-0002-8673-8254}
\author{Zhao Liu}
\email{liuz18@mails.tsinghua.edu.cn}
\orcid{0000-0002-5259-4504}
\affiliation{%
  \institution{Tsinghua University}
  \city{Beijing}
  \country{China}
}

\author{Jifa Zhang}
\email{jfzhang@zju.edu.cn}
\orcid{0000-0003-4761-7190}
\author{Tingwei Ji}
\email{zjjtw@zju.edu.cn}
\orcid{0000-0001-7556-6867}
\affiliation{%
  \institution{Zhejiang University}
  \city{Hangzhou}
  \country{China}
}

\author{Fangfang Xie}
\email{fangfang_xie@zju.edu.cn}
\orcid{0000-0001-5208-6086}
\affiliation{%
  \institution{Zhejiang University}
  \city{Hangzhou}
  \country{China}
}

\author{Xiaojing Lv}
\email{jing3704@126.com}
\orcid{0000-0003-1557-7550}
\affiliation{%
  \institution{National Supercomputing Center in Wuxi}
  \city{Wuxi}
  \country{China}
}

\author{Hanyue Liu}
\email{1424039642@qq.com}
\orcid{0009-0009-5724-2821}
\author{Xu Liu}
\email{aj_qlss@163.com}
\orcid{0000-0003-0886-8835}
\affiliation{%
  \institution{National Supercomputing Center in Wuxi}
  \city{Wuxi}
  \country{China}
}

\author{Xiyang Liu}
\email{17367765682@163.com}
\orcid{0009-0005-0230-3106}
\author{Xiaoyu Song}
\email{songxiaoyu33@163.com}
\orcid{0009-0004-7955-1522}
\affiliation{%
  \institution{Taiyuan University of Technology}
  \city{Taiyuan}
  \country{China}
}

\author{Guocheng Tao}
\email{guochengtao@intl.zju.edu.cn}
\orcid{0000-0003-2733-8039}
\affiliation{%
  \institution{Zhejiang University}
  \city{Hangzhou}
  \country{China}
}

\author{Yan Yan}
\email{Yan.Yan@xjtlu.edu.cn}
\orcid{0000-0003-0680-5806}
\affiliation{%
  \institution{Xi’an Jiaotong-Liverpool University}
  \city{Suzhou}
  \country{China}
}

\author{Paul Tucker}
\email{pgt23@cam.ac.uk}
\orcid{0000-0002-0874-3269}
\affiliation{%
  \institution{University of Cambridge}
  \city{Cambridge}
  \country{United Kingdom}
}
\author{Steven Miller}
\email{saem@ufl.edu}
\orcid{0000-0002-3697-3037}
\affiliation{%
  \institution{University of Florida}
  \city{Gainesville}
  \country{United States}
}
\author{Shirui Luo}
\email{shirui@illinois.edu}
\orcid{0000-0002-9360-1299}
\affiliation{%
  \institution{University of Illinois}
  \city{Urbana}
  \country{United States}
}
\author{Seid Koric}
\email{koric@illinois.edu}
\orcid{0000-0002-7330-6401}
\affiliation{%
  \institution{University of Illinois}
  \city{Urbana}
  \country{United States}
}

\author{Weimin Zheng}
\email{zwm-dcs@tsinghua.edu.cn}
\orcid{0009-0008-3026-2278}
\affiliation{%
  \institution{Tsinghua University}
  \city{Beijing}
  \country{China}
}
\renewcommand{\shortauthors}{Y. Fu, W. Shen, J. Cui, Y. Zheng, G. Yang et al.}

\begin{abstract}
A state-of-the-art large eddy simulation code has been developed to solve compressible flows in turbomachinery. The code has been engineered with a high degree of scalability, enabling it to effectively leverage the many-core architecture of the new Sunway system. 
A consistent performance of 115.8 DP-PFLOPs has been achieved on a high-pressure turbine cascade consisting of over 1.69 billion mesh elements and 865 billion Degree of Freedoms (DOFs).
By leveraging a high-order unstructured solver and its portability to large heterogeneous parallel systems, we have progressed towards solving the grand challenge problem outlined by NASA \cite{slotnick2014cfd}, which involves a time-dependent simulation of a complete engine, incorporating all the aerodynamic and heat transfer components.
\end{abstract}





\maketitle

\section{JUSTIFICATION FOR GORDON BELL PRIZE}

\begin{itemize}
  \item High complexity: Aero-thermodynamics of high pressure turbine blades with geometrical features across multiple orders of magnitude are modelled using an unstructured Computational Fluid Dynamics (CFD) code.
  \item High precision: The unstructured CFD code solves Navier-Stokes equations with numerical order of accuracy upto $8^{th}$ order.
  \item High fidelity: The numerical results are validated with detailed experimental results.
  \item High efficiency: A sustained peak performance of 115.8 DP-PFLOPs was achieved with strong scaling parallel efficiency measured at 89\%.
\end{itemize}

\section{PERFORMANCE ATTRIBUTES}

\begin{table}[H]
\centering
\caption{Performance attributes on Sunway supercomputer.}
\begin{tabularx}{\linewidth}{ll}
\toprule
Attributes & Values \\ 
\midrule
Category of achievement  & Scalability and peak  performance      \\
Performance  &  115.8 DP-PFLOPs     \\
Maximum problem size  & $ 865 $ billion DOFs   \\
Type of method used  & Explicit flux reconstruction \\
Results reported on basis of  & Application with and  without I/O  \\
Precision reported  & Double    \\
System scale & 350K MPEs with  22.4M CPEs\\
Measurement mechanism & FLOP count and timer \\
\bottomrule
\end{tabularx}
\label{table:perfomancetable}
\end{table}

\begin{figure*}[h]
    \centering
    \includegraphics[width=0.8\textwidth]{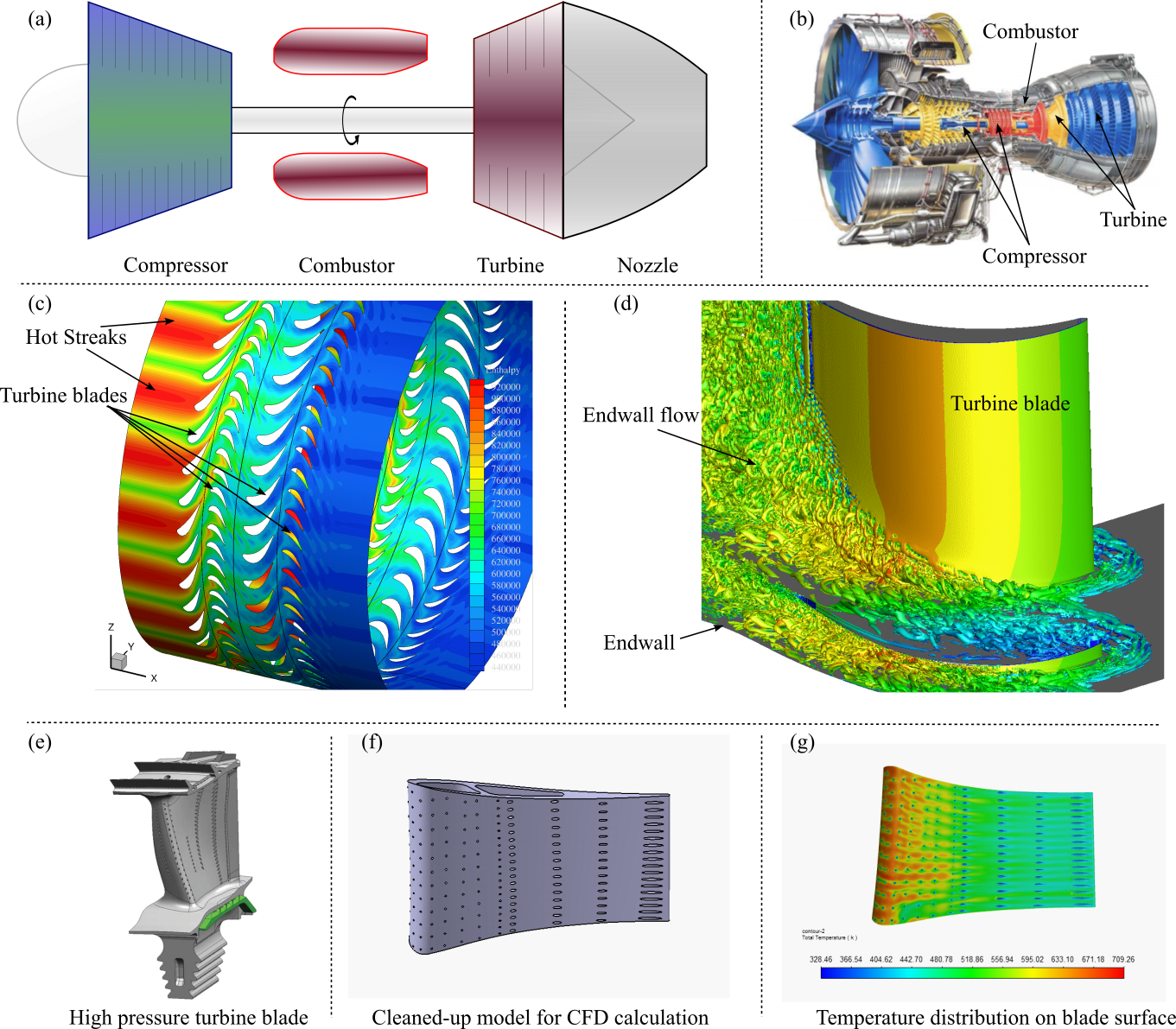}
    \caption{Main components of gas turbine and turbomachinery flow features: (a) schematic of core components of a gas turbine; (b)  illustration of a turbo-fan engine; (c) enthalpy field of hot streaks incoming from combustor into the high pressure turbine; (d) turbulent flow structures near the turbine endwall; (e) CAD model of a real turbine blade; (f) clean-up CAD model of a turbine blade with cooling holes for CFD calculation; (g) temperature field on high pressure turbine blade surface with cooling holes.}
    \label{fig:gt_diagram_hpt}
\end{figure*}

\section{OVERVIEW OF THE PROBLEM}
Turbomachinery remains a significant consumer of fossil fuels worldwide. At current fossil fuel prices, any small gas turbine performance improvement can result in a multi-billion-dollar economic impact and a major $CO_2$ emission reduction to mitigate the effects of global climate change. 
The bladed components of gas turbines, commonly referred as Turbomachines, are typically exposed to extremely high temperature and pressure, especially for the turbines located downstream of the combustor (i.e. high pressure turbine as illustrated in Fig. \ref{fig:gt_diagram_hpt}).
Given that the thermodynamic efficiency of a gas turbine is proportional to both the turbine entry temperature and compressor pressure ratio, there is a strong motivation to increase both parameters. 
However, this results in an elevated operating temperature (often above the material melting point) for high-pressure turbines. To mitigate the problem associated with high temperatures and pressures, intricate flow paths within blades are designed to channel cooling gas directly from the compressor outlet (as depicted in Fig. \ref{fig:gt_diagram_hpt}(e)). Reliable prediction of the blade temperature field is critical, as an error of just 2\% in predicting the blade metal temperature can halve its operating lifespan \cite{han2012gas}. Currently, a conservative design approach is typically employed to maintain the material and structural integrity from thermal damage. A higher fidelity prediction method could yield more accurate results. 
For instance, a 1\% reduction in cooling flow via a better thermal design could increase the overall efficiency of a gas turbine by 0.4\% \cite{han2012gas}. This could result in significant fuel savings worth billions of dollars, considering the number of gas turbines currently in operation \cite{sandberg2022fluid}.

The high pressure turbine blade is widely considered to be one of the most challenging components to simulate and design in gas turbines. Due to the highly unstable flow and temperature at the combustor exit, particularly under off-design conditions, heat transfer coefficient prediction errors can exceed 100\% \cite{gourdain2012rans}. The richness and complexity of the flow behavior within high pressure turbines contribute to the difficulty  of prediction. The internal cooling passages are long, narrow channels with widths on the millimeter scale and lengths on the centimeter scale. This wide range of scales, coupled with the complex geometry, presents a significant challenge in meshing the computational domain \cite{zheng2003novel, liou2003novel}.

Advanced CFD methods, such as Large Eddy Simulation (LES) or Direct Numerical Simulation (DNS), have been proposed to predict the flow and temperature of high pressure turbines with great accuracy \cite{tyacke2015future}.
Although these methods offer accurate and reliable simulation, they are computationally demanding. The cost of applying such high-fidelity methods is significantly higher compared to Reynolds-Averaged based methods. However, with the continuous advancement of high-performance computing power over the last decades, exascale computing is now within reach. By leveraging powerful computers and advanced numerical methods, more complex physics can be accurately captured and ultimately revealed to the designer. 

To take full advantage of the computing power of exascale clusters, a novel code called ZJU-FR (ZFR) has been developed from scratch to ensure portability across heterogeneous systems. Developed initially at the University of Florida \cite{Shen2020} and subsequently at Zhejiang University in China, ZFR is equipped with advanced turbomachinery flow capabilities, including rotating periodic boundary, sliding plane method, etc. 
The present study investigates the flow surrounding representative high pressure turbine blades from von K\'arm\'an institute \cite{arts1990aero} and NASA Lewis Research Center \cite{timko1984energy}.

\section{CURRENT STATE OF THE ART}
\subsection{Current State of Large Eddy Simulation}

Reynolds-Averaged Navier-Stokes (RANS) simulations, which are widely used in industrial and academic applications, possess several drawbacks, including the loss of temporal information and the lack of universality of the models. Despite the inherent complexity and unsteadiness of flows in gas turbines, turbomachinery designers still rely heavily on RANS simulation for day-to-day analysis. The circumferential averaging performed by RANS simulation between each blade row results in the smoothing out many important flow behaviors, such as wake-blade interaction. It has been demonstrated that these flow features are important in determining compressor and turbine efficiency, particularly for multi-stage components with many blade rows \cite{sandberg2022fluid, cui2016numerical}. 

As supercomputers are more accessible to research communities now, especially with the development of computation-focused devices like Graphics Processing Unit (GPU) or many-core devices, previously unaffordable DNS and LES are becoming more feasible for industrial applications. Researchers have used AVBP, a compressible Navier-Stokes solver developed by CERFACS, to conduct a full engine LES and investigate the complex flow features within gas turbines \cite{LES1, LES2}. The finite element method (FEM) based AVBP solver utilizes a Taylor-Galerkin discretization method to overcome the numerical instability of traditional FEM for convection equations. 
As part of the European Programming Environment for Heterogeneous Supercomputers (EPEEC) project, more advanced multi-layer parallel programming model, such as GASPI+MPSs and ArgoDSM+MPSs \cite{epeec}, have been integrated with AVBP to take advantage of the upcoming heterogeneous supercomputers.

\subsection{Current State of Numerical Methods and Related Solvers}

Computational mesh must be fine enough for LES to resolve the majority of turbulent scales and the time step size must be small enough to capture time-accurate flow dynamics, resulting in a computational cost orders of magnitude higher compared with RANS simulation \cite{GOPALAN2013249}. To address this challenge, it is essential to take advantage of the large modern massive parallel supercomputers. Almost all flagship systems feature a high FLOP/s-per byte ratio, meaning that memory speed is increasingly becoming the bottleneck of the computation. To keep up with the fast advancement of computing facilities, new series of finite-element-based high-order numerical schemes with higher data locality have been proposed, such as Discontinuous Galerkin (DG) method \cite{hesthaven2007nodal}, and the Flux Reconstruction (FR) method \cite{huynh2007flux}. These methods are hybrid structured/unstructured, with structurally aligned solution/flux points within each element while data associated with elements is unstructurally stored. This hybrid method offers flexibility when meshing complex domains, and fully utilizes the computing power of modern architectures due to its compactness and structuredness. 

A few finite-element-based high-order CFD solvers have been developed over the past few decades. NEK5000 \cite{nak5000} and Alya \cite{vazquez2016alya} served as a pioneering work that inspired and initiated the effort to push finite element methods for more practical industrial flows. Later, solvers based on DG method, such as SU2 \cite{Economon2016} and Flexi \cite{hindenlang2012explicit},  were developed to enable the solution of high-speed compressible flows on unstructured grids. A more unified framework called FR method or Correction Procedure via Reconstruction (CPR), was introduced later by Huynh in 2007 \cite{huynh2007flux} . The FR method can recover DG and spectral Difference (SD) methods \cite{SD} and is simpler and more computationally efficient than its predecessors \cite{Huynh2014}.

PyFR is an open-source framework based on the FR approach to solve advection diffusion type problems. It was developed at Imperial College London \cite{Witherden2014}, and is primarily written in Python. Mako template language is embedded in the code to generate numerical scheme kernel code for different hardware platforms at runtime, enabling developers to achieve high performance while still maintain a clean code base. The performance of FR methods is heavily limited by memory bandwidth on modern hardware. To alleviate this issue, a data structure facilitating cache blocking has been implemented, leading to a speedup of up to 2.8X for the Tyalor Green vortex test case \cite{Akkurt2022}. A petascale DNS simulation of compressible flow over a low pressure turbine (LPT) at Reynolds number of 200,000 was conducted. Meshes with up to 180 million hexahedron elements with fifth order accuracy were used, resulting in over 11 billion degree of freedoms per equation.
The result was selected as one of the finalists for the 2016 Gordon Bell prize.

\begin{figure}[htb]
    \centering
    \includegraphics[width=0.45\textwidth]{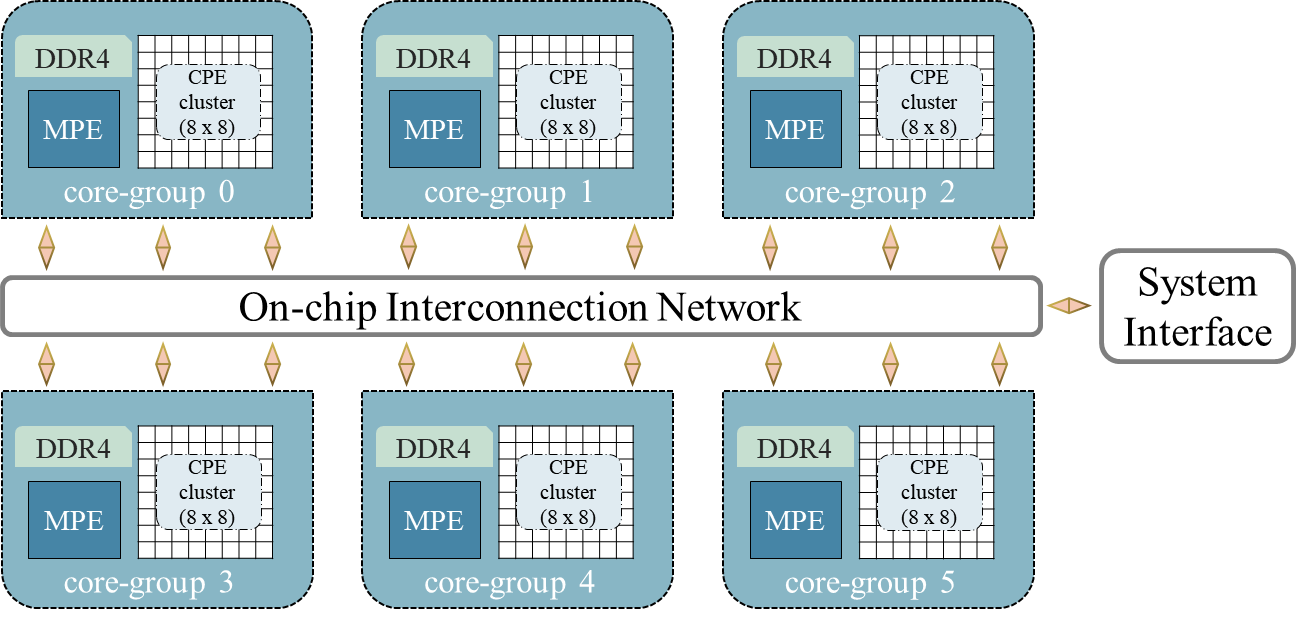}
    \caption{The architecture of SW26010Pro many-core processor on the new Sunway supercomputer.}
    \label{fig:SW20610pro}
\end{figure}

\subsection{Current State of New Sunway Supercomputer}

\subsubsection{System Architecture} 
The latest generation of the Sunway supercomputer has far more computing power than its predecessor (Sunway TaihuLight) thanks to a larger number of chips and the improved high-performance heterogeneous many-core processor. The system consists of over 100,000 SW26010 pro chips. Compared to the previous generation chip used in the Sunway TaihuLight, the upgraded processor includes six core-groups (CGs), each containing one MPE and one 8×8 CPE cluster with 16 GB DDR4 sharing memory (see Fig. \ref{fig:SW20610pro}). Each CPE has a local data memory (LDM) extended to 256 KB and supports wider 512 bit SIMD operation. More details of this new Sunway system can be found in \cite{hu20222, shang2021extreme}.

Sunway Accelerate Computing Architecture (SACA) is an essential component of the latest Sunway supercomputer. The SACA contains a parallel accelerated computing platform and programming model tailored to Sunway many-core architecture. SACA provides three levels of parallel acceleration, which are task (MPI), thread (Athread) and data (SIMD), that maximize system performance. By utilizing Direct Memory Access (DMA) and register communication, programmers can explicitly and effectively control data exchange on the LDM, which is crucial to minimize the communication cost of CPEs.

\subsubsection{Parallelization Challenge on Sunway for ZFR}
The ZFR solver faced several challenges to be efficiently ported on the Sunway system. 
To fully utilise the computing processor on the heterogeneous system, a hybrid "MPI + Athread" strategy incorporating multi-level parallelism needs to be adopted. 
Additionally, as an unstructured solver, it is impossible to perform regular partition like stencil computations on CPE cores.
Though FR method can significantly reduce data communication with high orders, bandwidth can still be the bottleneck, as Sunway has a relative lower byte-to-flop ratio compared to other heterogeneous systems. Thus, attention needs to be paid to data locality.
As the mesh scales to billions of elements and the solver runs with tens of millions of cores, an efficient parallel mesh pre-processing is also required.
With these challenges in mind, many innovative methods have been implemented on the pre-processor and the solver.

\section{INNOVATIONS REALISED}

\subsection{ZFR Framework}
\begin{figure}[h]
    \centering
    \includegraphics[width=0.4\textwidth]{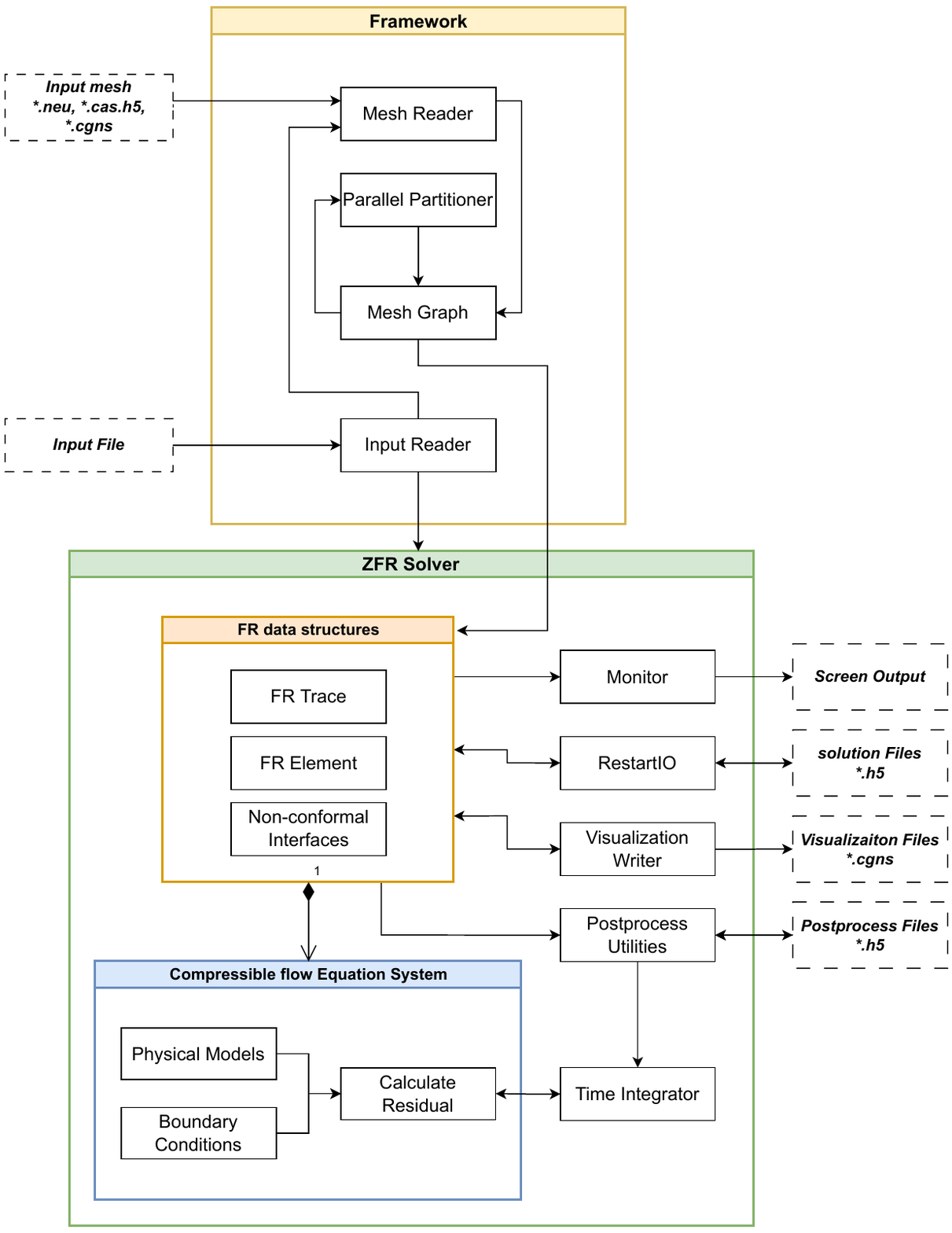}
    \caption{Overview of the code structure.}
    \label{fig:solver_struct}
\end{figure}

This section provides an overview of the structure of the inhouse code -- ZFR.
A schematic of the solver is illustrated in Fig. \ref{fig:solver_struct}.
The code contains two main components: a low-level supporting framework and a high-level FR solver.
The low-level framework incorporates algorithms and data structures that are independent of numerical schemes, offering an unified interface for high-performance physics simulation.
The parallel partitioner is used to decompose the computational domain into non-overlapping sub-domains for parallel computation, with either ParMETIS \cite{Padua2011} or PT-Scotch \cite{CHEVALIER2008318} working as the backend.

The FR solver comprises several components that define the geometrical data structures, numerical algorithms, and equation systems.
Only the equation system encapsulates the physical problem while other components are problem independent.
The current implementation supports the compressible flow equation system, with physical models including boundary conditions, Riemann solvers, body forcing, LES sub-grid scale models, and shock capturing methods.

\subsubsection{FR Algorithm}
The governing equations are rewritten into the form of conservation law and transformed to reference domain within each element as
\begin{equation}
    \frac{\partial \hat{\mathbf{Q}}}{\partial t}+\nabla_\xi\cdot(\hat{\mathbf{F}}+\hat{\mathbf{G}})=\hat{\mathbf{S}},
\end{equation}
where $\hat{\mathbf{Q}}(\boldsymbol{\xi},t)=|\mathbf{J}|\mathbf{Q}(x,t)$ is the transformed solution variable, $\mathbf{\hat{F}}=|\mathbf{J}|\mathbf{J}^{-1}\cdot \mathbf{F}$ and $\mathbf{\hat{G}}=|\mathbf{J}|\mathbf{J}^{-1}\cdot \mathbf{G}$ are the transformed inviscid and viscous flux, $\nabla_\xi\cdot\Box$ is divergence operator in the reference domain, and $\hat{\mathbf{S}}=|\mathbf{J}|\mathbf{S}$ is the transformed source term. Here,
$\mathbf{J}_{ij}=\frac{\partial \boldsymbol{x}_i}{\partial\boldsymbol{\xi}_j}
$
is the transformation Jacobian, where $\boldsymbol{x}$ represents coordinates in physics domain, and $\boldsymbol{\xi}$ represents the coordinates in reference domain.
The transformed solution and flux functions are approximated by piecewise discontinuous polynomials defined in the reference element as
\begin{subequations}
    \begin{align}
        \hat{\mathbf{Q}}^\delta(\boldsymbol{\xi}) & =\sum_{i=1}^{N_s}\hat{\boldsymbol{Q}}^\delta(\boldsymbol{\xi}_i) l_i(\boldsymbol{\xi}), \\
        \hat{\mathbf{F}}^\delta(\boldsymbol{\xi}) & =\sum_{i=1}^{N_s}\hat{\mathbf{F}}^\delta(\boldsymbol{\xi}_i) l_i(\boldsymbol{\xi}),     \\
        \hat{\mathbf{G}}^\delta(\boldsymbol{\xi}) & =\sum_{i=1}^{N_s}\hat{\mathbf{G}}^\delta(\boldsymbol{\xi}_i) l_i(\boldsymbol{\xi}),
    \end{align}
\end{subequations}
where $N_s$ is number of solution points inside each element.
On the interfaces, the normal common numerical flux is calculated using an approximate Riemann solver with solutions interpolated to both sides of the interface.
Then, a correction function $g$ is used to correct the divergence of flux with the normal common numerical flux on the interface to recover a global solution.
In this work, $g$ is chosen so that the resulting scheme is equivalent to a nodal collocated DG method \cite{huynh2007flux}.
The viscous flux is handled using the local discontinuous Galerkin (LDG) method, which is described in detail by Huynh \cite{Huynh2009}.
The semi-discretized form of the scheme takes the form
\begin{multline}
    \frac{\text{d}\mathbf{Q}}{\text{d} t}= \\
    -|\mathbf{J}|^{-1}\left[\nabla_\xi\cdot\mathbf{\hat{F}}^\delta+\sum_{i=1}^{N_I}\sum_{j=1}^{N_{Fi}}\nabla_\xi\cdot g_{ij}(\mathbf{\boldsymbol{\xi}})\left(\mathbf{\hat{F}}^{\delta I\bot}_{ij}-\mathbf{\hat{F}}^{\delta \bot}_{ij}\right)\right] \\
    +\mathbf{S},
\end{multline}
where $N_I$ and $N_{Fi}$ are the number of interfaces on each element and number of flux points on the interface, respectively.
The transformed numerical flux, and the normal transformed discontinuous flux on the interface are represented by $\mathbf{\hat{F}}^{\delta I\bot}$ and $\mathbf{\hat{F}}^{\delta \bot}$, respectively.
Finally, the solution is advanced in time using an ODE solver such as strong-stability-preserving Runge-Kutta method \cite{ketcheson2008highly}.
\subsubsection{Pre-processor Parallelization}
The preprocess module of the code is carefully designed to efficiently read and process large scale computational mesh in parallel.
Mesh entities including cells and vertices to be read by each rank are determined using a cumulative storage format, so that each rank reads a non-overlapping chunk of data.
For CGNS file format, this operation is performed using the parallel I/O support of HDF5 built on MPI-IO, either collectively or independently.

After reading the cell and vertex information, the mesh partitioner is called to construct a dual-graph, which includes cell-cell connectivity information.
The face list is constructed on each rank using local cell connectivity, and the vertex list of each face is sorted in ascending order to create a unique numbering.
The face list is then sorted in ascending order by vertex IDs to make sure that coupled faces are next to each other.
Then, local internal faces are coupled and repeating entries are eliminated, so that the remaining uncoupled faces are either MPI internal faces or boundary faces.

To read the boundary faces, algorithm \ref{alg:bf_read}  presents the pseudo code that uses the cumulative storage format to read each boundary.
According to the CGNS specification, each boundary condition corresponds to one or more face element sections.

\begin{algorithm}
\caption{Method to read boundary faces with CGNS.}\label{alg:bf_read}
\begin{algorithmic}[1]
\Procedure{ReadBoundary}{boundaryList,faceList}
\State $\text{boundaryList} \gets \text{read boundary condition info}$
\State $\text{glob\_bface} \gets \text{get global number of boundary faces}$
\State $\text{read\_range} \gets \textbf{DistributeEntity}(\text{glob\_bface})$
\For{$ibnd \gets 1$ to $N$}
\State $\text{boco\_sect} \gets \text{get section list}$ 
\For{$isec \in \text{boco\_sect}$}
\State $\text{section\_range} \gets \text{get section range}$
\If{$\text{read\_range}\cap \text{section\_range}\neq \emptyset$}
\State $\text{vlist} \gets \text{read connectivity}$
\State $\text{update bface\_list with vlist and }ibnd$
\EndIf
\EndFor
\EndFor

\State distribute uncoupled faces with leading vertex ID
\State $\text{sbuffer} \gets \text{uncoupled faces}$

\State $np \gets \text{number of processes}$
\State \textbf{MPI\_Issend}(\text{sbuffer,}$p$) for $p\gets 1,np$
\Repeat
\If{\textbf{MPI\_Iprobe}}
\State \textbf{MPI\_Recv}(\text{rbuffer})
\State update face\_recv with rbuffer
\EndIf
\If{\textbf{MPI\_Issend} finished}
\State \text{call} \textbf{MPI\_Ibarrier}
\EndIf
\Until{\textbf{MPI\_Ibarrier} completed}

\State \textbf{sort}(face\_recv)
\State remove repeating entries in face\_recv
\State distribute boundary faces with leading vertex ID
\State $\text{sbuffer} \gets \text{boundary faces}$

\State \textbf{MPI\_Issend}(\text{sbuffer,}$p$) for $p\gets 1,np$
\Repeat
\If{\textbf{MPI\_Iprobe}}
\State \textbf{MPI\_Recv}(\text{rbuffer})
\State update bface\_recv with rbuffer
\EndIf
\If{\textbf{MPI\_Issend} finished}
\State \text{call} \textbf{MPI\_Ibarrier}
\EndIf
\Until{\textbf{MPI\_Ibarrier} completed}

\State \textbf{sort}(bface\_recv)
\State match bface\_recv with face\_recv
\State $\text{sbuffer} \gets \text{matched faces}$

\State \textbf{MPI\_Issend}(\text{sbuffer,}$p$) for $p\gets 1,np$
\Repeat
\If{\textbf{MPI\_Iprobe}}
\State \textbf{MPI\_Recv}(\text{rbuffer})
\State update faceList with rbuffer
\State update boundaryList with rbuffer
\EndIf
\If{\textbf{MPI\_Issend} finished}
\State \text{call} \textbf{MPI\_Ibarrier}
\EndIf
\Until{\textbf{MPI\_Ibarrier} completed}
\EndProcedure
\end{algorithmic}
\end{algorithm}

MPI internal faces in each rank need to find their remote counterparts.
The face matching information is used later in the solver to exchange data within each time step.
The  algorithm to find MPI matching faces is presented in Algorithm \ref{alg:mpi_face_match}.
Uncoupled MPI internal faces are redistributed linearly throughout the ranks by their leading vertex ID.
This method makes sure that coupled faces will be sent to the same rank.
Similar to the local internal face matching algorithm, the received faces are sorted and their counterparts are found.

\begin{algorithm}
\caption{Method to match MPI internal faces.}\label{alg:mpi_face_match}
\begin{algorithmic}[1]
\Procedure{MatchMPIFaces}{MPIFaceRank,faceList}
\State distribute uncoupled faces with leading vertex ID
\State $\text{sbuffer} \gets \text{uncoupled faces}$
\State $np \gets \text{number of processes}$
\State \textbf{MPI\_Issend}(\text{sbuffer,}$p$) for $p\gets 1,np$
\Repeat
\If{\textbf{MPI\_Iprobe}}
\State \textbf{MPI\_Recv}(\text{rbuffer})
\State update face\_recv with rbuffer
\EndIf
\If{\textbf{MPI\_Issend} finished}
\State  \text{call} \textbf{MPI\_Ibarrier}
\EndIf
\Until{\textbf{MPI\_Ibarrier} completed}
\State \textbf{sort}(face\_recv)
\State match face\_recv
\State $\text{sbuffer} \gets \text{coupled faces}$
\State \textbf{MPI\_Issend}(\text{sbuffer,}$p$) for $p\gets 1,np$
\Repeat
\If{\textbf{MPI\_Iprobe}}
\State \textbf{MPI\_Recv}(\text{rbuffer})
\EndIf
\If{\textbf{MPI\_Issend} finished}
\State  \text{call} \textbf{MPI\_Ibarrier}
\EndIf
\Until{\textbf{MPI\_Ibarrier} completed}
\State update faceList with rbuffer
\State update MPIFaceRank with rbuffer
\EndProcedure
\end{algorithmic}
\end{algorithm}

\subsection{Three-level Parallelization Design}

\begin{figure}[h]
    \centering
    \includegraphics[width=0.5\textwidth]{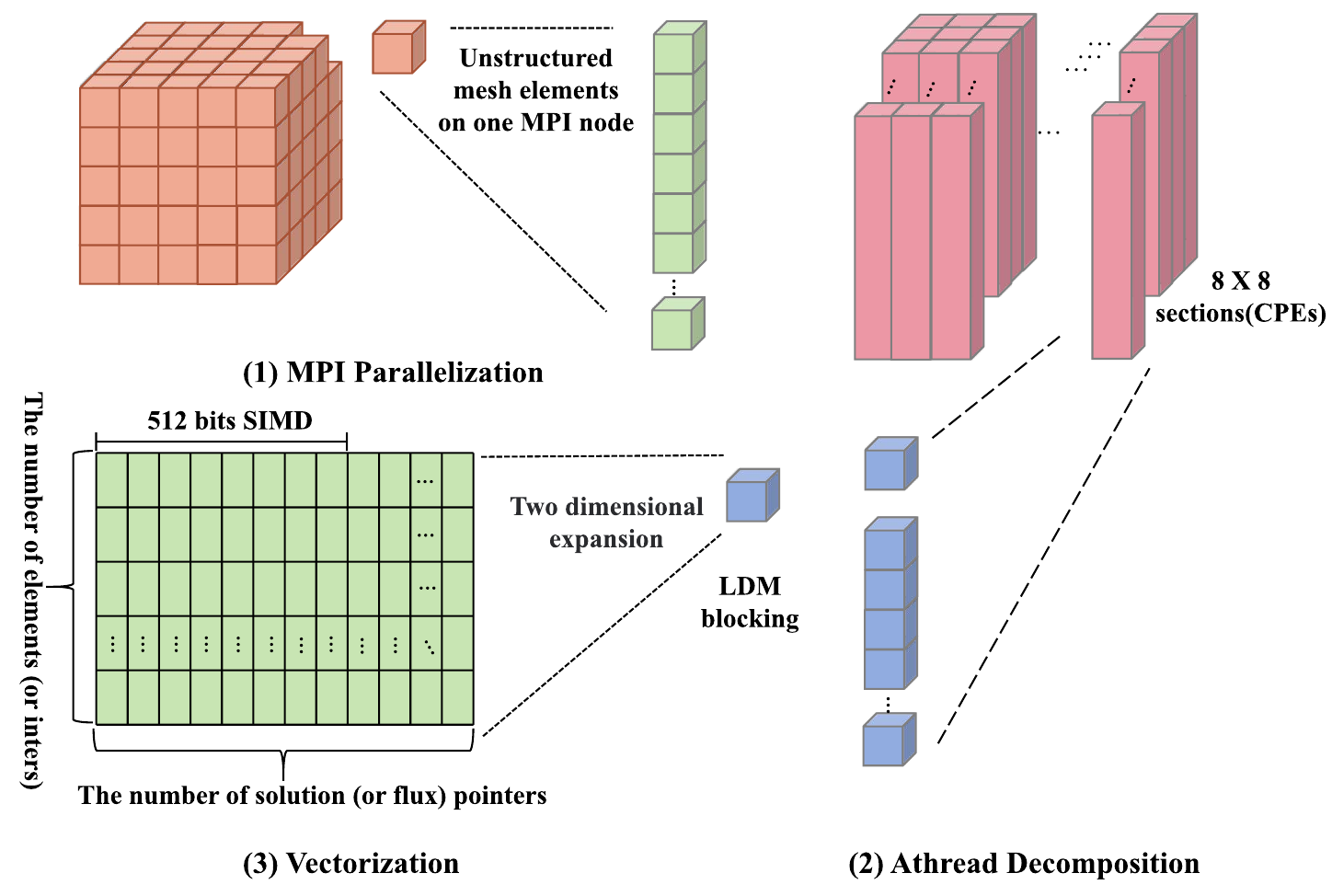}
    \caption{The muti-level parallelization design.}
    \label{fig:parallel}
\end{figure}

To optimize computational performance on the Sunway system, we have implemented a three-level parallelization strategy using the SACA programming model. This approach, as illustrated in Fig. \ref{fig:parallel}, involves decomposing the unstructured grid into different partitions for each MPI process, distributing computing tasks evenly across elements and interfaces for each CPE thread, and conducting vectorization on the data dimension of sequential points per element or interface. The success of the last two levels of parallelism relies on maximizing utilization of the limited LDM space on CPEs. To achieve these, we have implemented the following three levels of parallelizatioin.

\subsubsection{MPI Parallelization}
The ZFR solver utilizes a decomposition strategy to partition the entirety of an unstructured grid into multiple sub-computation domains. Each domain is assigned a partition weight that is based on the type of element it contains, aiming to optimize the load balance across MPI processes.

\subsubsection{LDM Blocking for CPE Thread}
The SW26010 Pro’s memory hierarchy presents a simple approach to second-level parallelism by equally distributing computational workload of the entire sub-domain across 64 CPE threads, while LDM operates as a cache with 256B cache lines. This design is similar to traditional multi-cores processor. However, it can lead to significant performance degradation due to long data access latency and cache miss.
To improve the efficiency of CPEs’ access to main memory and avoid complex data layout design caused by cache lines, we employed a LDM blocking technique and managed the LDM explicitly, inspired by cache blocking. 
Our method involves selecting a consistent quantity of computation data (e.g., elements or interfaces) by local data id, partitioning computation data into small blocks along the order of memory address, and processing each block in a sub-loop of thread. 
After blocking, the size of each block is limited by the 256 KB LDM space and size of unit data. 
The data transfer between CPEs is highly efficient by leveraging the DMA channel.
However, when launching point-wise kernels with indirect memory access (PI) which is unavoidable in many unstructured solvers, a noticeable portion of data block can not be filled along the memory address directly, resulting in a landslide in computing performance. 
To address this issue, we analyzed the layout of variables and found out address continuity of the lowest level of data (n\_fpts\_per\_inter) on one interface. We then packed them as the basic unit of data communication by DMA. In computationally intensive kernels, we applied a double-buffering approach to reuse the LDM data and hide the memory latency. Based on the LDM blocking strategy, data blocks are divided into two smaller ones, allowing the computation of the current block while a CPE thread is prefetching the next block data in a pipeline way.

\subsubsection{Vectorization}
Vectorization is an crucial step in parallel acceleration to achieve efficient utilization of the chip’s performance. While the SACA programming model provides automatic vectorization and optimization schemes, the ZFR solver did not achieve the expected speedup after enabling this feature. To address this issue, we opted to perform an 8-double width loop tile on the lowest level of data layout and refactor a series of related subroutines with SIMD.

\subsection{Point-wise Kernels Fusion Scheme}
On the many-core architecture, the performance of FR method is significantly limited by the memory bandwidth. While we have employed various strategies, such as LDM blocking to maximize bandwidth utilization, double buffering in some kernels to hide variables communication, and shared data between CPEs through register communication to alleviate limited LDM space, memory access still dominates almost all of the point-wise kernel runtime after vectorization. Through analysis of these point-wise kernels, we found that a dataset transferred from LDM by one point-wise kernel is latter consumed by another. With this understanding, we propose a point-wise kernel-fusion scheme to reduce redundant memory bandwidth expenditure.

\begin{figure}[h]
    \centering
    \includegraphics[width=0.45\textwidth]{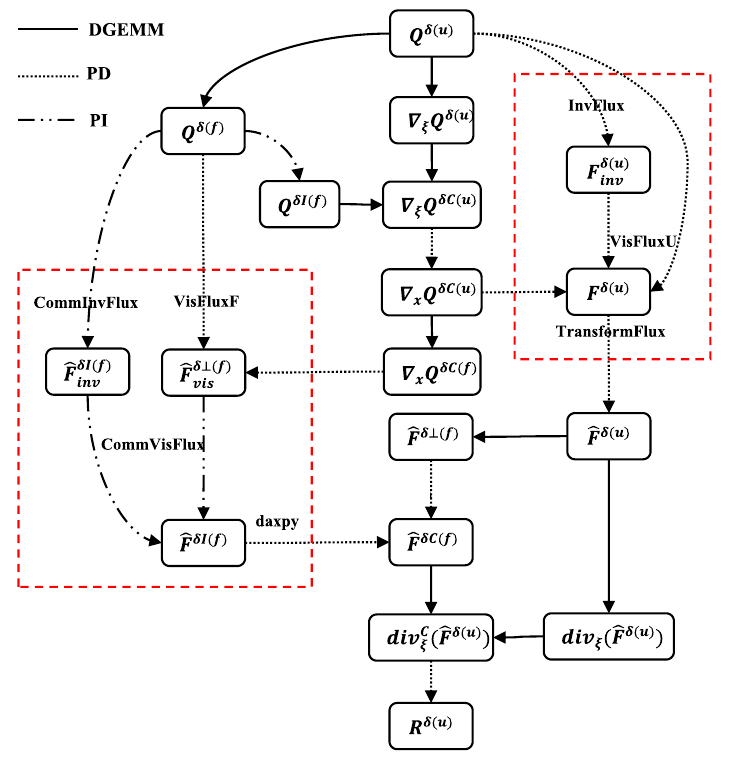}
    \caption{DAG for data dependencies.}
    \label{fig:kernelFusion1}
\end{figure}

To describe the details of the fusion scheme, we first introduce a directed acyclic graph (DAG) that indicates the data dependencies. Fig. \ref{fig:kernelFusion1} shows the entire structure of the DAG, which includes nodes representing the main variable in the ZFR solver and edges determined by kernels representing the direction of the data flow. 
Three line-types were used to denote the different kernels categories: matrix multiplications (DGEMM), point-wise kernels with direct memory access patterns (PD), and point-wise kernels with some level of indirect memory access(PI) \cite{Akkurt2022}. The DAG provides a clear overview of the entire data flow and dependencies. 

\begin{figure}[h]
    \centering
    \includegraphics[width=0.45\textwidth]{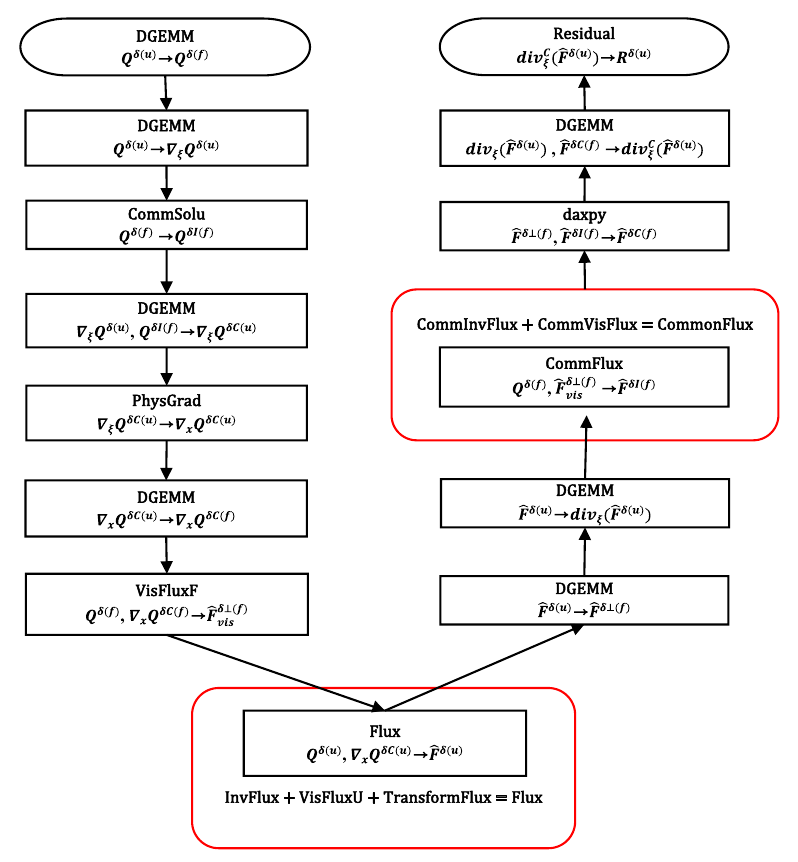}
    \caption{Point-wise kernels fusion scheme.}
    \label{fig:kernelFusion}
\end{figure}

The optimal kernel fusion is to consolidate all data onto a single node on the DAG and perform all computations using that kernel.
However, with xMath \cite{Liu2022} as the third-party library for high performance matrix multiplication, it is impossible to achieve a full kernel fusion without the original code of the library.
In this work, we primarily focuses on PI and PD kernels for the fusion and find out two potential fusion region on the DAG, which is framed by red dotted line. 
We successfully fused five kernels into two optimized ones, as shown in detail in Fig. \ref{fig:kernelFusion}.
 
\section{HOW PERFORMANCE WAS MEASURED}

\subsection{Test Case Design}
The performance of the code are measured using two cases: the LS89 cascade and the vane of the first stage high pressure turbine of GE-E$^3$ \cite{timko1984energy}. The LS89 cascade is relatively simpler with no cooling holes. However, it has more detailed measuremnt data. Therefore, scalling test and code validation are performed using the LS89 cascade. The GE-E$^3$ test case, on the other hand, is more representative to the real turbine blade with cooling holes and complex internal channels.

\subsubsection{Test Case 1}
\begin{figure}[h]
    \centering
    \includegraphics[width=0.45\textwidth]{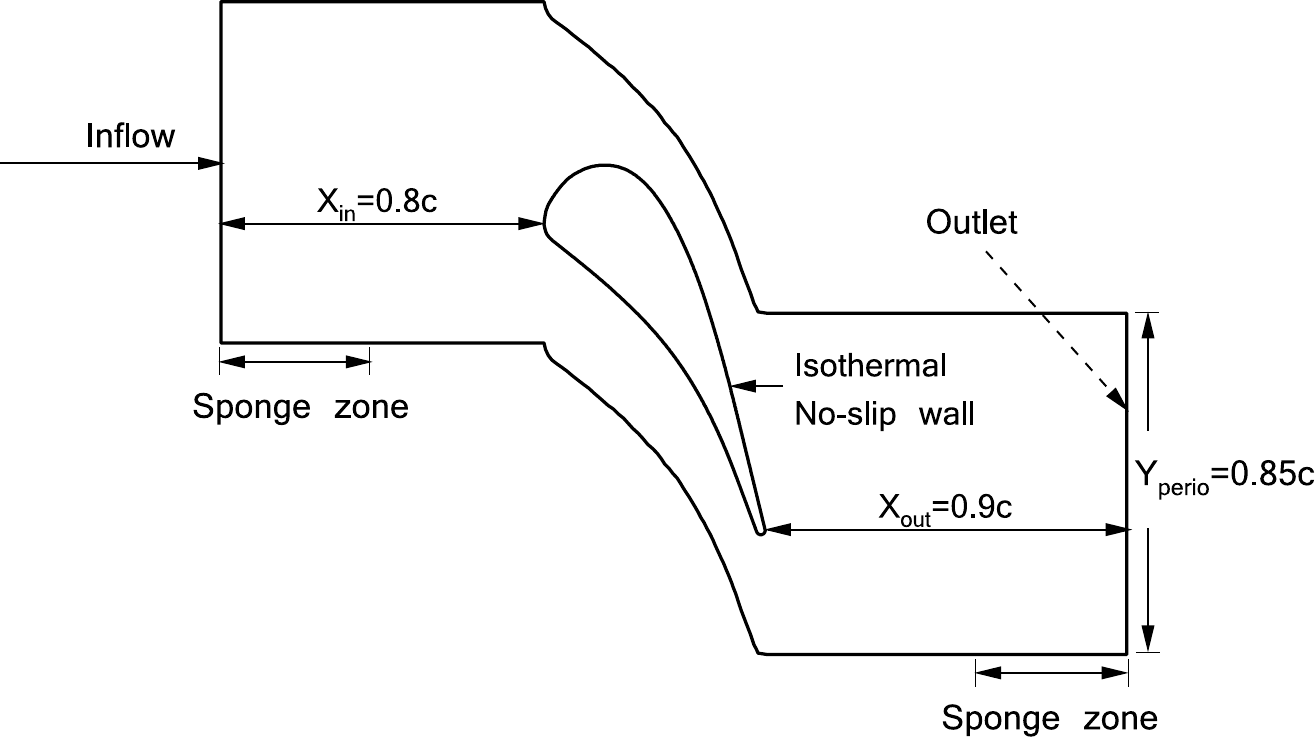}
    \caption{Computational domain of the high pressure turbine testcase.}
    \label{fig:geometry}
\end{figure}

Performance measurement and accuracy validation are conducted by solving the flow of the LS89 cascade. The test conditions and geometry are based on MUR129 experimental data obtained at the von K\'arm\'an Institute \cite{arts1990aero}. The case specifics are summarized in Table \ref{table:testcase}. The computational domain with the adopted boundary conditions are shown in Fig. \ref{fig:geometry}. It is noted that two sponge zones are imposed near the inlet and outlet boundaries to damp out any fluctuations and prevent reflections from boundaries. 

\begin{table}[h!]
\centering
\caption{Test Case 1 specifics.}
\begin{tabularx}{0.5\linewidth}{ll} 
\toprule
Parameter & Value \\ 
\midrule
Chord (C) & $ 67.647 $mm      \\
Pitch  &  $ 0.85 $C     \\
Stagger angle  & $ 55.0^{\circ} $   \\
$ Mach_{exit} $  & $ 0.84 $      \\
$ Mach_{inlet} $ & $ 0.15 $    \\
Reynolds Number & $ 0.57 \times 10^6 $ \\
\bottomrule
\end{tabularx}
\label{table:testcase}
\end{table}

A total of five sets of unstructured meshes were generated, with the element counts given in Table \ref{table:mesh}. Mesh1 was employed to validate the framework and case setup. This validation run enabled the time averaging, sponge zones, statistics collection, and parallel data output with fourth-order solution polynomials interpolation within each element. Mesh2 to Mesh5 were designed to run the weak scaling test, Mesh2 and Mesh3 were also used for the strong scaling test. The peak sustained performance was measured using the Mesh5. 

\begin{table}[h!]
\centering
\caption{Mesh element counts of Test Case 1.}
\begin{tabularx}{0.6\linewidth}{lc} 
\toprule
Mesh & Mesh elements counts (million) \\ 
\midrule
Mesh1  & 1.7          \\
Mesh2  & 211         \\
Mesh3  & 422         \\
Mesh4  & 844         \\
Mesh5  & 1,689      \\
\bottomrule
\end{tabularx}
\label{table:mesh}
\end{table}

\begin{figure}[h]
    \centering
    \includegraphics[width=0.4\textwidth]{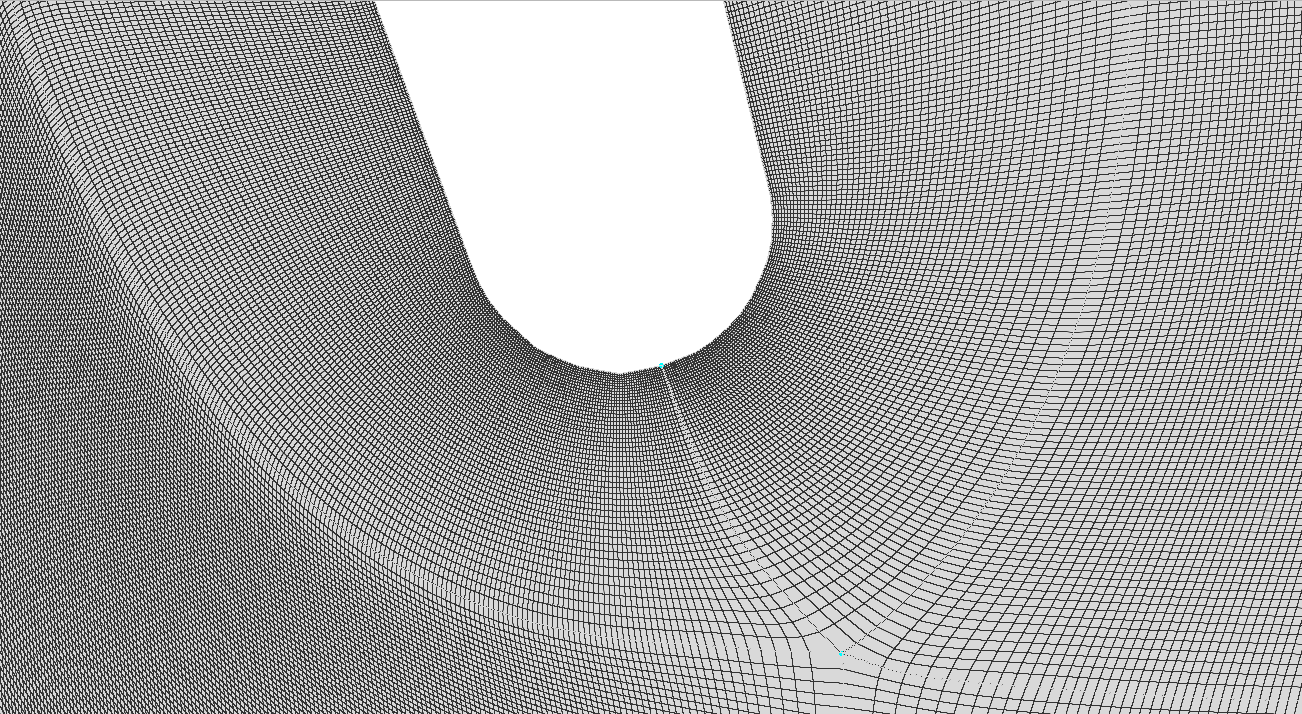}
    \caption{Cross-sectional view of the mesh around the blade trailling edge.}
    \label{fig:mesh}
\end{figure}

\subsubsection{Test Case 2}

\begin{figure}[h]
    \centering
\includegraphics[width=0.45\textwidth]{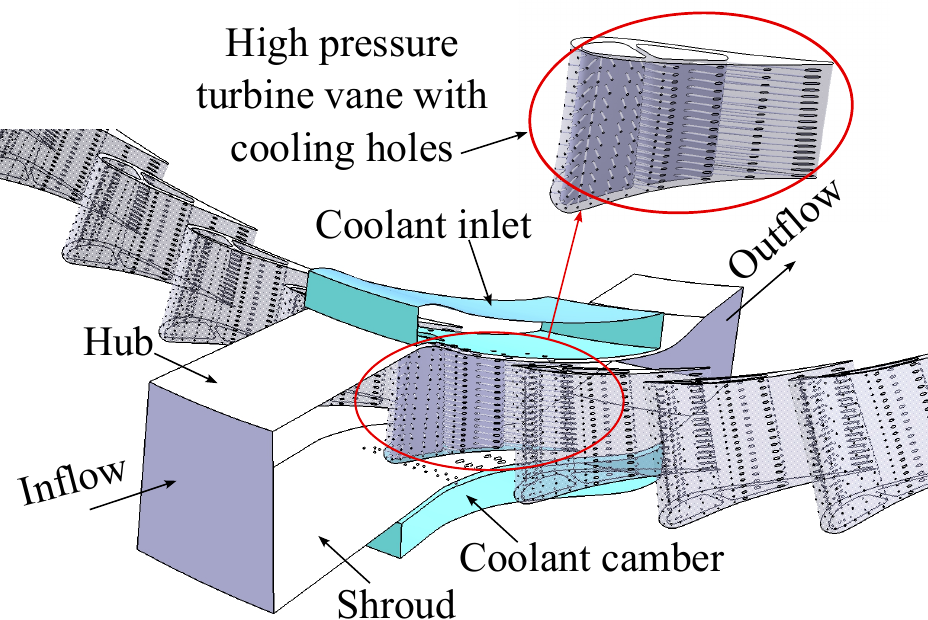}
    \caption{Geometrical model and boundary condition of high pressure turbine vane with cooling holes.}
    \label{fig:geometry2}
\end{figure}

To demonstrate the current solver can simulate complex practical cases, flow within the vane of the first stage high pressure turbine of GE-E$^3$ \cite{timko1984energy} is calculated. 
Fig. \ref{fig:geometry2} shows the three-dimensional model of the turbine vane and the boundary condition implemented in the current study. With periodic boundary condition applied in the circumferential direction, one flow passage is solved. Relatively low temperature coolant is injected into the computational domain near the hub. The coolant channels into the internal passage of the turbine vane, and then ejects to the mainstream from the small cooling holes. The detailed boundary condition and mesh element counts can be found in Table \ref{table:testcase2}.

\begin{table}[h!]
\centering
\caption{Test Case 2 boundary condition and mesh element counts.}
\begin{tabularx}{0.8\linewidth}{ll} 
\toprule
Parameter & Value \\ 
\midrule
Total temperature (main inlet)  & $ 709 $K     \\
Total temperature (coolant inlet)  & $ 399 $K     \\
Total pressure (main inlet)  & $ 3.4474 \times 10^5$ Pa     \\
Total pressure (coolant inlet)  & $ 3.5095 \times 10^5$ Pa     \\
Pressure ratio (inlet total to outlet static)  & $ 1.63 $     \\
$ Mach_{inlet} $  & $ 0.1 $   \\
Mesh element counts & $ 10 $ million \\
Degree of freedom & $ 640 $ million \\
\bottomrule
\end{tabularx}
\label{table:testcase2}
\end{table}

\begin{figure}[h]
    \centering
\includegraphics[width=0.45\textwidth]{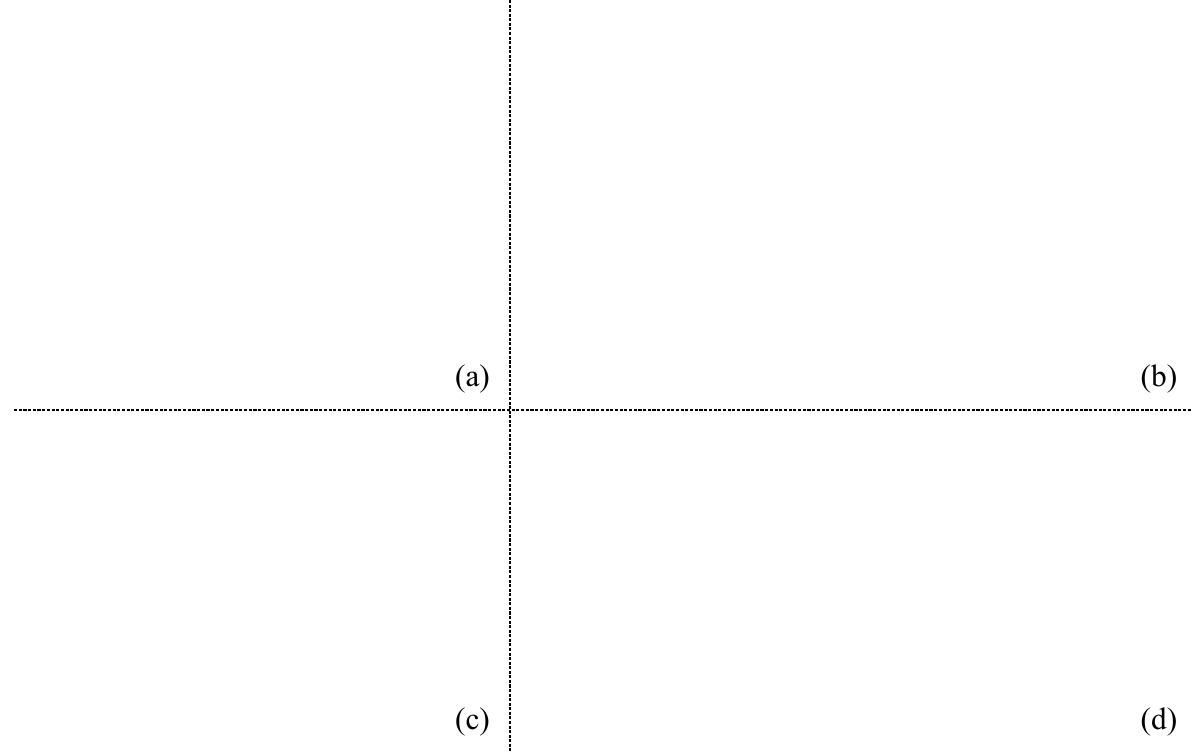}
    \caption{Surface mesh of the high pressure turbine vane on (a) internal channels and cooling holes and (b) the junction of leading edge and endwall and (c) the pressure surface and (d) the trailing edge.}
    \label{fig:mesh2}
\end{figure}

\subsection{FLOPs Measurement}
All numerical tests were performed on the new Sunway supercomputer using double precision. The performance tests were designed to evaluate the scaling and efficiency of the solver with no post-processing module enabled (i.e. no I/O or statistical data collection). The performance is measured by averaging time taken for one time step, after running a test for 50 time steps. The total number of floating point operations (FLOPs) of one time step is estimated in a manner similar to that in \cite{PyfrBG}. The CPE kernels were divided into two groups: matrix multiplications and point-wise calculation. In matrix-multiplication kernels, the floating point operations was counted as 2mnk, with m representing the number of rows in A and C matrix, n representing the number of columns in B and C matrix, k representing the common dimension of A and B matrix. This estimate was less than the actual FLOPs executed in the ZFR solver, resulting in a conservative estimation of the overall FLOPs. For point-wise kernels, we used two methods to count the number of operations. The first method counted all double-precision instructions in the assembly code, while the other counted the specific algorithm and operation of the numerical scheme. Both methods produced similar operation counts. Here, we chose the latter for the final number.

\section{PERFORMANCE RESULTS}

\subsection{Scaling tests for Test Case 1}
\subsubsection{Strong scaling}
\begin{figure}[h]
    \centering
    \includegraphics[width=0.42\textwidth]{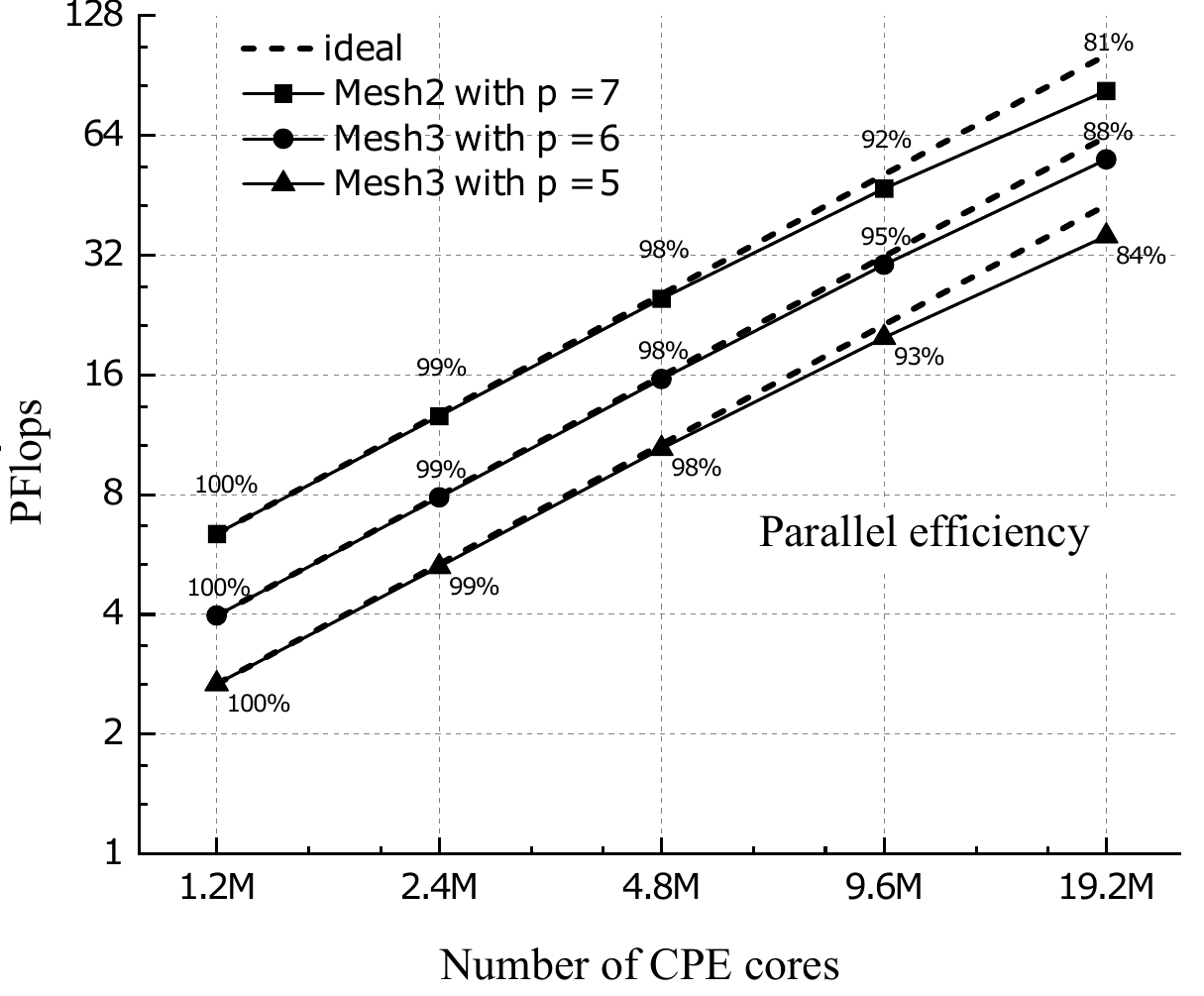}
    \caption{Strong scaling of ZFR on the latest generation Sunway supercomputer for different meshes and polynomial orders.}
    \label{fig:strongscaling}
\end{figure}

Fig. \ref{fig:strongscaling} shows the strong scaling test results on the latest generation Sunway supercomputer. The 211 million and 422 million unstructured meshes of LS89 case were benchmarked and achieved the excellent parallel efficiency from 1.2 million to 19.2 million CPE cores. At 19.2 million CPE cores, the parallel efficiency of three configurations all dropped to below 90$\%$. This is due to the fact that each CPE core is only allocated with 11 elements or 22 elements for Mesh2 and Mesh3 respectively, at which the CPE core cannot be fully loaded. Considering the configuration of Mesh3 with different polynomial orders, it is observed that the efficiency of order p = 6 is higher than that with p = 5. The strong scalling test result suggests that the element size on each CPE plays a vital role in the parallel performance.

\subsubsection{Weak scaling}
\begin{figure}[h]
    \centering
    \includegraphics[width=0.42\textwidth]{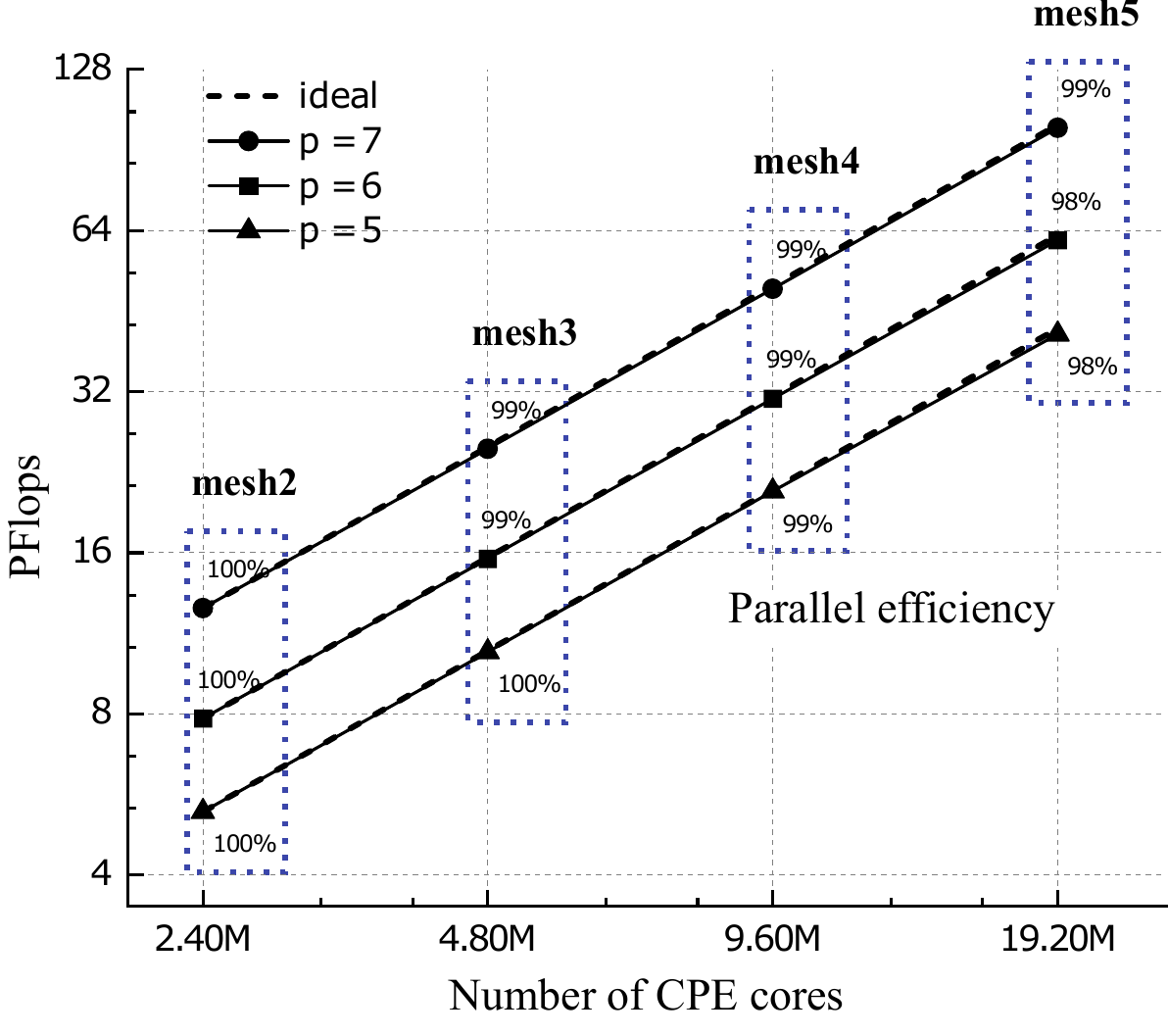}
    \caption{Weak scaling of ZFR on the latest generation Sunway supercomputer for different polynomial orders.}
    \label{fig:weakscaling}
\end{figure}

For weak scaling, four sets of meshes (see Table \ref{table:mesh}) have been generated, with element counts ranging from 211 million to 1.69 billion. 
Fig. \ref{fig:weakscaling} shows the weak scaling test result, where 2.4 million to 19.2 million CPE cores are used in the test. Excellent weak scaling performance is observed up to 19.2 million CPE cores with various orders.

\subsection{Peak sustained performance for Test Case 1}
The peak performance was benchmarked using Mesh5 with a seventh order of polynomial, resulting in 1.69 billion elements and 865 billion DOFs. The case was ported to 60\% of the Sunway machine with 350 thousand MPE cores and 22.4 million CPE cores, achieving a sustained performance of 115.8 DP-PFLOPs.

\subsection{Physics revealed through simulation with high complexity, high fidelity, high precision, and high efficiency}
\subsubsection{Test Case 1} 
The physics run was performed on Mesh1 with a third order polynomial. 
The chosen Courant-Friedrichs-Lewy (CFL) value was 1, requiring approximately 100,000 time steps to complete one flow pass over the blade chord. 
The case commenced with a $1^\text{st}$ order of accuracy to flush out transients, and then switched to the fourth order for a high fidelity run.
Following the low order run, a total of 300,000 time steps was initiated to achieve a fully developed state, after which time averaging was performed over another 300,000 time steps.
The entire simulation was performed on 2000 nodes with 768,000 CPEs for 40 wall clock hours.

\begin{figure}[h]
    \centering
    \includegraphics[width=0.42\textwidth]{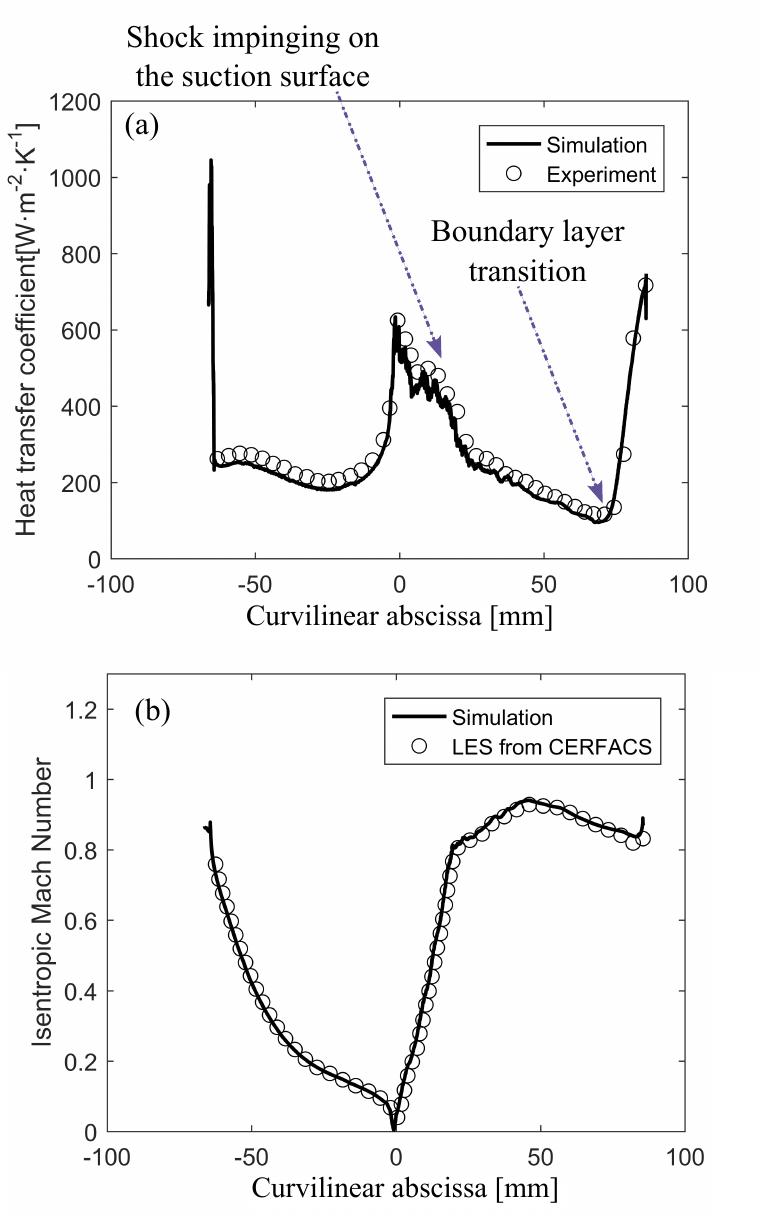}
    \caption{Comparison between numerical results and  measurements.}
    \label{fig:validation}
\end{figure}

\begin{figure}[h]
    \centering
    \includegraphics[width=0.4\textwidth]{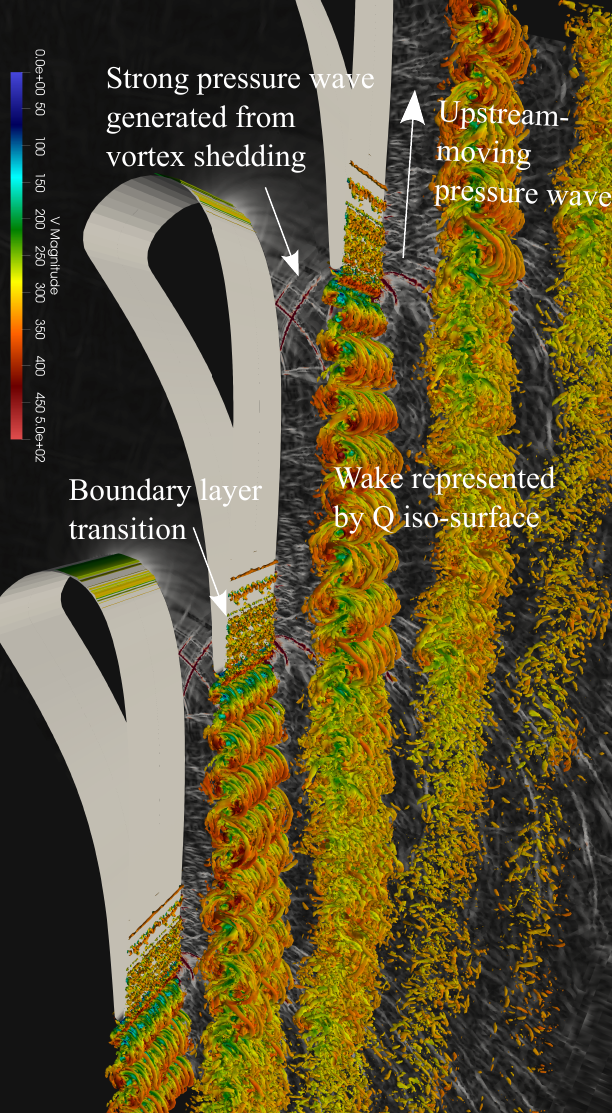}
    \caption{The screenshot of Q-criteria iso-surface colored with velocity magnitude, with background contoured by pressure gradient to visualize the shock wave and pressure propagation.}
    \label{fig:snapshot}
\end{figure}

Fig. \ref{fig:validation} compares the simulation results with experimental data. It is noted that numerical results compare favourably with experimental and numerical results in literature for both heat transfer coefficient and isentropic Mach number over the blade surface. The convective heat transfer coefficient is notoriously difficult to predict with errors of over 50\% for RANS simulations \cite{tyacke2015future}. The fact that the current high fidelity simulation can accurately predict the heat transfer coefficient is that the rich flow physics is captured. Fig. \ref{fig:snapshot} shows a snapshot of the instantaneous flow with turbulent flow structures visualized by Q-criterion \cite{Dubief2000} and shock wave and wake propagation by pressure gradient in the background. The vortex shedded from the blade trailing edge generates strong pressure waves, which impinge on the neighbouring blade suction surface, creating a region with oscillating heat transfer coefficient (see fig. \ref{fig:validation}). On the suction surface near trailing edge, boundary layer transition triggered by the upstream moving pressure wave is also evident, which is consistent with the sharp increase of heat transfer coefficient near the trailing edge.

\subsubsection{Test Case 2}

\begin{figure}[h]
    \centering
\includegraphics[width=0.42\textwidth]{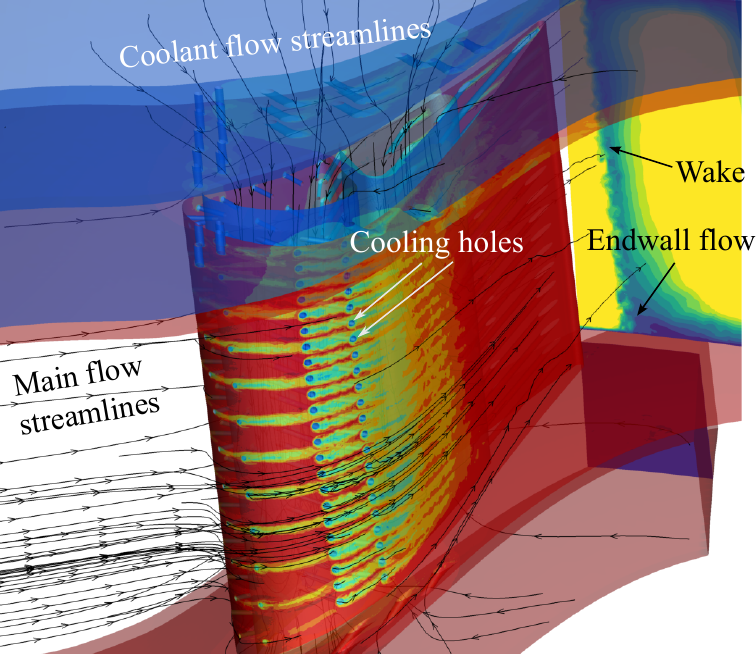}
    \caption{Temperature over the suction surface of the high pressure turbine vane with coolant ejecting from cooling holes.}
    \label{fig:GE-flow}
\end{figure}

Only physics runs were performed for Test Case 2. Similar case running strategy as Test Case 1 was used. The case started with 1st order of accuracy and then switched to the fourth order to resolve smaller turbulent scales. Fig. \ref{fig:GE-flow} shows the instantaneous temperature field over the blade suction surface with main and coolant flow streamlines embedded. It is evident that the temperature of the blade surface is lowered significantly by the coolant ejecting from the cooling holes. Different to the Test Cases 1, hub and shroud are included in Test Case 2, resulting in the evident endwall flow downstream of the trailing edge. Though there is no surface blade temperature and heat transfer coefficient measurement for this test case, it is believed that this high fidelity simulation can confidently predict the blade surface temperature as more physics are captured, such as jet ejecting to the mainstream, endwall flows, wake shedding from the trailing edge, etc.

\section{IMPLICATIONS}

The present study details the development of a LES code, which is fully portable to the new Sunway architecture. The code includes a parallel pre-processor that can handle large meshes with billions of elements and efficiently find matching faces for MPI communication. The code was validated and profiled using high-pressure turbine testcases, and numerical results compared favorably with experimental data, particularly for the challenging prediction of heat transfer coefficient. The ability to accurately predict heat transfer is crucial for predicting blade temperature and material integrity in industrial turbomachinery applications, particularly for transitional and separated flows. With higher confidence in simulation, more boundary-pushing and groundbreaking designs can be made to increase overall engine efficiency and eventually reduce $CO_2$.
Our work demonstrates the potential of high order FR methods and heterogeneous computing power for LES of turbomachinery flows. 

\subsection{Parallelization Optimization for FR method on Heterogeneous Computing Architecture}

We have demonstrated in this work that the Sunway architecture can be efficiently utilized for high-order CFD simulations on unstructured meshes, through the implementation of multilevel parallelism technique. Our results demonstrate excellent scalabilty up to 300,000 MPI ranks with 19.2 million CPE cores.
Furthermore, while the Sunway architecture differs from other heterogeneous platforms in terms of its memory model, we anticipate that such multi-level parallelism technique can be employed to optimize algorithms for other platforms as well.
With support from other technologies, such as adaptive mesh generation with high-order elements, the flux reconstruction method can be a competitive candidate for the future whole engine simulation.

\subsection{Large Eddy Simulation for Turbomachinery Flows}
Nowadays, CFD plays an increasingly important role in turbomachinery's aerodynamic, thermal, and aeromechanic design, resulting in shorter design cycles, safer and more efficient performance, and reduced weight and costs.
However, RANS based turbulence model often underestimates the diffusion-driven turbulent mixing. This affects the prediction of almost every turbomachinery component.
LES, on the other hand, offers substantially more accurate and detailed information about the flow field than RANS, but requires drastically more computational resources. In our research, we demonstrated how LES can accurately model sizable sections of practical turbomachinery blades on the emerging exascale computing technology. Since the scale of computing used in cutting-edge HPC systems, like Sunway, is about 10-15 years ahead of that used by the leading industrial adopter \cite{top500}, this pioneering research will open the door for future high-fidelity and complex numerical studies for turbomachinery flows, encouraging more innovative designs with short turnaround time to achieve safer and more efficient gas turbines with lower emissions.

\begin{acks}
We acknowledge the support from the National Nature Science Foundation of China with grant number 52106060 and 92152202. The opensourced HiFiLES and SU2 project at Stanford University and the PyFR project at Imperial College London inspired some of our algorithm development in this work, which are also acknowledged here. We also thank Will Trojak at IBM for useful technical discussions.
\end{acks}

\printbibliography

\end{document}